\documentclass[conference]{IEEEtran}
\IEEEoverridecommandlockouts
\usepackage[latin1]{inputenc}
\usepackage{amsmath,amssymb,amsfonts}
\usepackage{bm}
\usepackage{cite}
\usepackage{graphicx}
\usepackage{microtype}
\usepackage{flushend}
\usepackage{stfloats}
\usepackage{multirow}
\usepackage{xcolor}
\usepackage{url}
\usepackage[colorlinks=true,linkcolor=blue,citecolor=blue,urlcolor=blue,anchorcolor=blue]{hyperref}
\usepackage[figure]{hypcap}

\AddToHookNext{shipout/foreground}{%
  \put(0,-10mm){%
    \makebox[\paperwidth][c]{%
      \parbox[t]{\textwidth}{\centering\normalfont\fontsize{10}{10}\selectfont
        This work has been submitted to the IEEE for possible publication. \\Copyright may be transferred without notice, after which this version may no longer be accessible.}%
    }%
  }%
}

\begin{document}
\bstctlcite{IEEErefcompact}

\title{Decision-Focused Learning via Neural Optimization Layers for Battery-Assisted Grid-Import Smoothing in AI Data Centers
}

\author{\IEEEauthorblockN{Xinda Zheng\textsuperscript{1}, Canchen Jiang\textsuperscript{1*}, Frits de Nijs\textsuperscript{1}, Edward Lam\textsuperscript{1}, Markus Wagner\textsuperscript{1}}
\IEEEauthorblockA{\textsuperscript{1} Department of Data Science and AI, Faculty of Information Technology, Monash University, Australia \\
}
\thanks{*Corresponding author: Canchen Jiang (canchen.jiang1@monash.edu).}
}

\maketitle

\begin{abstract}
Rapid AI data-center power fluctuations challenge grid integration. Sustained grid-import smoothing requires allocating limited battery power and energy under uncertain demand. We formulate risk-aware BESS scheduling that coordinates advance reserve commitments and demand-responsive dispatch, accounting for operating costs, grid-ramp penalties, and their conditional value-at-risk. We develop battery scheduling decision-focused learning (BS-DFL) with a constraint-aware neural optimization layer trained directly from this objective without optimal-solution labels. The trained layer is then frozen to guide demand-predictor learning through scheduling costs and ramping risk, enabling solver-free rolling-horizon scheduling. In a 10,000-minute simulation using loads derived from measured AI workloads, BS-DFL reduces mean absolute minute-to-minute grid-import changes by 71.70\% and total scheduling cost by 18.01\% relative to an MSE-trained predictor coupled with a mixed-integer linear programming solver, with an average decision time of 18.77 ms. 
\end{abstract}

\section{Introduction}
The rapid growth of AI data centers introduces highly variable electricity loads, creating challenges for grid-compatible operation and power-ramp management \cite{ko2026oscillations}. Behind-the-meter battery energy storage systems (BESSs) can mitigate rapid load fluctuations without modifying computational workload schedules \cite{ross2026gridforming}. However, sustained grid-import smoothing is constrained by battery power and energy limits, requiring careful scheduling of charging and discharging flexibility. Predictive scheduling is therefore needed to mitigate current grid-import fluctuations while preserving sufficient power and energy headroom for future uncertain demand. 

Existing studies have developed battery-assisted ramp-rate control for AI data centers~\cite{abera2026ramp}. At the scheduling level, coordinated workload shifting, battery dispatch, and reserve allocation have been investigated to balance operating costs and grid-ramping risk~\cite{xie2027risk}. To guide advance battery scheduling, neural load forecasts have been combined with robust model predictive control for peak shaving~\cite{mary2025robust}, while probabilistic load forecasts support capacity allocation between peak shaving and frequency containment reserves~\cite{henni2022industrial}. Nevertheless, prediction errors of similar magnitude can lead to substantially different scheduling costs and grid-ramp violations, depending on the battery's available power and energy flexibility.

Conventional forecasting approaches typically minimize statistical prediction losses, such as mean squared error, without explicitly considering their impact on downstream battery scheduling decisions. This motivates decision-focused learning (DFL), which incorporates downstream scheduling costs into predictor training~\cite{zhang2026decision}. DFL has been applied to BESS reserve-market participation~\cite{paredes2025reserve}, PV-battery scheduling~\cite{depoortere2026pvbattery}, and AI data-center workload scheduling under distribution-grid constraints~\cite{yan2026pts}. These methods rely on differentiable convex optimization layers, limiting their direct application to scheduling with discrete decisions. To accommodate such decisions, surrogate-loss and black-box differentiation methods have been applied to battery scheduling~\cite{alkhulaifi2026decision}. Other approaches include perturbation-based learning for energy storage arbitrage~\cite{yi2025perturbed} and supervised neural optimizer for battery peak shaving and energy cost reduction~\cite{kim2025neural}. However, these approaches often rely on optimization solvers during training or deployment, which can limit training scalability for large or complex problems and the responsiveness of real-time battery scheduling.

More recently, ~\cite{zheng_dflnol_manuscript} trains a neural optimization layer (NOL) directly from optimization objectives and uses it to guide predictor learning, enabling solver-free training and deployment. This framework offers a way to align prediction with downstream decision while supporting real-time decision making. Nevertheless, the existing NOL architecture is not directly applicable to BESS scheduling, which involves intertemporal energy constraints, charging/discharging exclusivity, and uncertain demand. To address these challenges, we propose a battery scheduling decision-focused learning framework (BS-DFL), which uses a BESS-scheduling-specific NOL (BS-NOL) to train demand predictors for improved battery scheduling and generate feasible decisions for the mixed-integer formulation without optimization solvers during training and rolling-horizon deployment. The main contributions are summarized as follows:

\begin{itemize}
\setlength{\itemsep}{0pt}
\setlength{\parsep}{0pt}
    \item We formulate a risk-aware BESS scheduling framework for AI data centers that coordinates advance reserve commitments and demand-responsive dispatch, accounting for operating costs, grid-import ramp penalties, and their tail risks through CVaR. 
    \item We develop BS-DFL with BS-NOL that combines scenario-aware encoding and sequential constraint-aware decoding to coordinate reserve commitments and demand-responsive dispatch. The layer is trained directly using the scheduling objective without optimal solution labels, then frozen to guide demand-predictor learning through downstream scheduling costs and ramping risks.
    \item Using facility load profiles generated from measured AI workloads, we demonstrate that BS-DFL reduces mean absolute grid-import changes by 71.70\% and total operating costs by 18.01\% relative to an MSE-trained predictor coupled with the Gurobi solver. It achieves a total cost within 4.39\% of the perfect-information rolling-horizon reference while computing decisions in milliseconds.
\end{itemize}

\section{System Model and Problem Formulation}\label{sec:problem_formulation}
We consider an AI data center with a behind-the-meter BESS connected to the grid at a POI. The BESS is operated using a rolling-horizon two-stage stochastic scheduling framework to coordinate economic operation and mitigation of rapid POI-import variations. 

The horizon $\mathcal T=\{1,\ldots,T\}$ consists of one-minute intervals of duration $\Delta t$. A finite scenario set $\mathcal S$ represents demand uncertainty, with interval-average demand $\widehat P_{t,s}^{\mathrm{DC}}$ and probability $\rho_s\ge0$, where $\sum_{s\in\mathcal S}\rho_s=1$. Scenarios combine demand forecasts with historical variations (Section~\ref{sec:dfl_framework}).

\subsection{Stage 1: BESS Scheduling and Reserve Allocation} \label{stage1}
Let $b_t$ denote the baseline net BESS power, with positive and negative values indicating discharging and charging, respectively. The upward and downward reserve power capacities are denoted by $q_t^+$ and $q_t^-$, satisfying
\begin{subequations}\label{eq:reserve_limits}
\begin{align}
-P_{\max}^{\mathrm{ch}} &\le b_t \le P_{\max}^{\mathrm{dis}},\\
0\le q_t^{+} &\le P_{\max}^{\mathrm{dis}}-b_t,\\
0\le q_t^{-} &\le P_{\max}^{\mathrm{ch}}+b_t,
\end{align}
\end{subequations}
where $P_{\max}^{\mathrm{ch}}$ and $P_{\max}^{\mathrm{dis}}$ denote the BESS rated charging and discharging power. The corresponding extreme net outputs are 
\begin{equation}\label{eq:extreme_power}
\begin{aligned}
& v_t^{+}=b_t + q_t^{+}, &
& v_t^{-}=b_t - q_t^{-}.
\end{aligned}
\end{equation}
To ensure that idle operation remains feasible within the committed power envelope, we impose
\begin{equation}\label{eq:idle_option}
v_t^{-} \le 0 \le v_t^{+}.
\end{equation}

\subsection{Stage 2: Scenario-Dependent BESS Operation}

For scenario $s$, let $a_{t,s}$ denote the corrective power activation. The net BESS power $p_{t,s}^{\mathrm B}$ and its charging/discharging components satisfy
\begin{subequations}\label{eq:bess_operation}
\begin{align}
& p_{t,s}^{\mathrm{dis}}-p_{t,s}^{\mathrm{ch}} =b_t+a_{t,s},\\
& -q_t^- \le a_{t,s}\le q_t^+,\\
& 0\le p_{t,s}^{\mathrm{dis}} \le z_{t,s}P_{\max}^{\mathrm{dis}},\\
& 0\le p_{t,s}^{\mathrm{ch}} \le(1-z_{t,s})P_{\max}^{\mathrm{ch}},
\end{align}
\end{subequations}
where $p_{t,s}^{\mathrm{ch}}$ and $p_{t,s}^{\mathrm{dis}}$ are the actual charging and discharging powers, and the binary variable $z_{t,s}\in\{0,1\}$ prevents simultaneous charging and discharging at time $t$ for scenario~$s$.

The stored battery energy $E_{t,s}$ evolves according to
\begin{subequations}\label{eq:soc_dynamics}
\begin{align}
& E_{t,s}=E_{t-1,s} +\eta_{\mathrm{ch}}p_{t,s}^{\mathrm{ch}}\Delta t -\frac{p_{t,s}^{\mathrm{dis}}\Delta t} {\eta_{\mathrm{dis}}}, \\
& E_{\min}\le E_{t,s}\le E_{\max}, 
\end{align}
\end{subequations}
where $\eta_{\mathrm{ch}}$ and $\eta_{\mathrm{dis}}$ denote charging and discharging efficiencies. The initial energy is fixed at $E_{0,s}=E_0$.

To sustain either committed power limit for one interval, the starting energy must provide sufficient stored energy and charging headroom:
\begin{subequations}\label{eq:reserve_energy}
\begin{align}
E_{t-1,s}&\ge E_{\min}+\frac{\Delta t}{\eta_{\mathrm{dis}}}v_t^+,\\
E_{t-1,s}&\le E_{\max}+\eta_{\mathrm{ch}}\Delta t\,v_t^-.
\end{align}
\end{subequations}

\subsection{POI Import and Ramping Constraints} \label{subsec:grid}
With grid export prohibited, the projected POI import $g_{t,s}\ge0$ under forecast scenario $s$ is
\begin{equation}\label{eq:grid_power}
g_{t,s}=\widehat{P}_{t,s}^{\mathrm{DC}} +p_{t,s}^{\mathrm{ch}} -p_{t,s}^{\mathrm{dis}}.
\end{equation}

To mitigate rapid fluctuations at the POI, upward and downward grid-power changes satisfy
\begin{equation}\label{eq:grid_ramp_limits}
-(R^\downarrow+\delta_{t,s}^-) \leq g_{t,s}-g_{t-1,s} \le R^\uparrow+\delta_{t,s}^+,
\end{equation}
where $R^\uparrow$ and $R^\downarrow$ are the specified upward and downward ramp limits, and $\delta_{t,s}^+,\delta_{t,s}^-\ge0$ measure the corresponding exceedances. Initially, $g_{0,s}=g_0$.
The accumulated ramp-violation cost under scenario $s$ is
\begin{equation}\label{eq:violation_cost}
V_s=\sum_{t\in\mathcal T} \left(\lambda^{+} \delta_{t,s}^{+} +\lambda^{-} \delta_{t,s}^{-} \right),
\end{equation}
where $\lambda^{+}$ and $\lambda^{-}$ represent penalties for upward and downward violations, respectively.

\subsection{Risk-Aware Stochastic Optimization}
The objective is to minimize reserve allocation costs, expected operating costs, and the risk of severe POI ramp violations.

The reserve allocation cost is defined as $C^{\mathrm{res}} =\sum_{t\in\mathcal T} (c_t^{+} q_t^{+} + c_t^{-} q_t^{-})\Delta t$, 
where $c_t^{+}$ and $c_t^{-}$ denote the incremental capacity-commitment cost coefficients for upward and downward reserves, respectively. These coefficients penalize excessive reserve allocation separately from the energy-arbitrage costs of BESS operation.

The scenario-dependent operating cost is $C_s^{\mathrm{op}} =\sum_{t\in\mathcal T} \left[\pi_t g_{t,s}+c^{\mathrm{deg}} \left(p_{t,s}^{\mathrm{ch}} +p_{t,s}^{\mathrm{dis}}\right) \right]\Delta t$, where $\pi_t$ denotes the electricity price and $c^{\mathrm{deg}}$ is a throughput-based battery degradation cost coefficient.

The terminal penalty $C_s^{\mathrm{term}}=c^{\mathrm{term}}|E_{T,s}-E^{\mathrm{tar}}|$ penalizes deviations from the target energy at the horizon's end.

Let $\mathbf x^{(1)}=\{b_t,q_t^+,q_t^-\}_{t\in\mathcal T}$ denote the first-stage decisions and let $\mathbf x^{\mathrm{rec}} = \{a_{t,s},p_{t,s}^{\mathrm{ch}},
p_{t,s}^{\mathrm{dis}},E_{t,s},z_{t,s},
g_{t,s},\delta_{t,s}^+,\delta_{t,s}^-\}_
{t\in\mathcal T,s\in\mathcal S}$ collect scenario-dependent operating variables, including the binary operating modes. Let $\mathcal F$ denote the feasible set defined by \eqref{eq:reserve_limits}-\eqref{eq:grid_ramp_limits}.

CVaR measures the tail risk of the scenario ramp cost $V_s$ in~\eqref{eq:violation_cost}. With auxiliary variables $\zeta\in\mathbb R$ and $\xi_s\ge0$, the optimization problem is
\begin{align} \label{eq:stochastic_model}
\min_{\mathbf x^{(1)},\mathbf x^{\mathrm{rec}},\zeta,\boldsymbol{\xi}} \quad &C^{\mathrm{res}} +\sum_{s\in\mathcal S}\rho_s \left( C_s^{\mathrm{op}}+V_s +C_s^{\mathrm{term}} \right)
\nonumber\\
&+\beta\left[ \zeta+\frac{1}{1-\alpha} \sum_{s\in\mathcal S}\rho_s\xi_s
\right] \\
\mathrm{s.t.}\quad
&(\mathbf x^{(1)},\mathbf x^{\mathrm{rec}})\in\mathcal F, \quad
\xi_s\ge V_s-\zeta,\quad\forall s. \notag 
\end{align}
where $ \alpha \in (0,1)$ denotes the CVaR confidence level and $\beta \ge 0$ controls the emphasis on tail risk.

\subsection{Real-Time Dispatch and Rolling-Horizon Implementation}
\label{subsec:rolling_operation}
At each decision epoch, only the first-interval baseline and reserves $(b_1^*,q_1^{+,*},q_1^{-,*})$ from \eqref{eq:stochastic_model} are committed. Once updated demand measurements become available, BESS dispatch is re-optimized over the operational horizon using the current demand estimate and updated future forecasts. The dispatch optimization minimizes the operating, ramping, and terminal costs defined above, subject to the BESS and POI constraints in \eqref{eq:bess_operation}-\eqref{eq:grid_ramp_limits}, with the first-interval baseline and reserves fixed at their committed values. Only the first-interval dispatch is implemented, while future decisions remain provisional. The battery state and POI import are updated, and the stochastic scheduling problem is re-solved for the next interval.

\section{Decision-Focused Learning Framework}\label{sec:dfl_framework}
\begin{figure}[!h]
\centering
\includegraphics[width=1\linewidth]{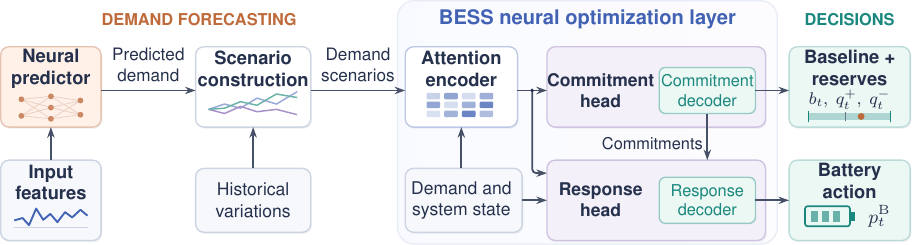}
\caption{BS-DFL framework. Forecast-based scenarios guide baseline and reserve commitments. Battery responses adapt to observed demand and state within the committed range. The two decoders convert network outputs into feasible scheduling decisions.}
\label{fig:bs_dfl_framework}
\end{figure}

BS-DFL combines a neural demand predictor with the constraint-aware BS-NOL, as shown in Fig.~\ref{fig:bs_dfl_framework}.

Given an input feature vector $\mathbf h$ containing historical demand, node occupancy, and time features, a neural predictor $f_\theta$ produces the positive demand forecast vector $\boldsymbol\mu_\theta=f_\theta(\mathbf h)=(\mu_{\theta,1},\ldots,\mu_{\theta,T})^\top$ over the planning horizon. To represent uncertainty around this forecast, we select historical demand profiles from different training days with similar recent load levels and variability, node occupancy, and time of day. Let $H_{t,s}$ denote the demand at interval $t$ in the $s$th selected historical demand profile. The deviations $d_{t,s}=H_{t,s}-\frac{1}{|\mathcal S|}\sum_{j\in\mathcal S}H_{t,j}$ from the mean of the selected profiles at each interval are combined with the forecast to form demand scenarios $\widehat P_{t,s}^{\mathrm{DC}}$ with probabilities $\rho_s=1/|\mathcal S|$:
\begin{equation}\label{eq:scenario_prediction}
\widehat P_{t,s}^{\mathrm{DC}}
=\operatorname{softplus}\!\left(\operatorname{softplus}^{-1}(\mu_{\theta,t})+d_{t,s}\right).
\end{equation}
This transformation keeps demand positive and returns the original forecast when $d_{t,s}=0$. BS-NOL generates first-stage commitments $\mathbf x^{(1)}$, defining the committed power range in~\eqref{eq:extreme_power}, and scenario-dependent operating decisions $\mathbf x^{\mathrm{rec}}$, including battery responses within that range as in~\eqref{eq:bess_operation}.

\subsection{Battery-Scheduling Neural Optimization Layer}\label{subsec:battery_nol}
The encoder processes demand values with their time and scenario
indices. Attention learns how to combine information across times
and scenarios into a shared summary, which is combined with the
initial battery state to guide reserve commitments and battery responses.

A linear commitment head maps the shared summary to
$\gamma_t^-,\gamma_t^+,\gamma_t^{\mathrm b}\in[0,1]$.
The commitment decoder uses these coefficients to select the lower
power limit, upper power limit, and baseline. From the starting energy levels, it first computes power bounds that satisfy the power and reserve-energy constraints in every scenario:
\begin{subequations}\label{eq:decoder_bounds}
\begin{align}
\ell_t&=\max\!\left\{-P_{\max}^{\mathrm{ch}},
\frac{\max_s E_{t-1,s}-E_{\max}}{\eta_{\mathrm{ch}}\Delta t}\right\},\label{eq:decoder_lower_bound}\\
u_t&=\min\!\left\{P_{\max}^{\mathrm{dis}},
\frac{\eta_{\mathrm{dis}}(\min_s E_{t-1,s}-E_{\min})}{\Delta t}\right\}.\label{eq:decoder_upper_bound}
\end{align}
\end{subequations}
Within these bounds, $\gamma_t^-$ selects the charging limit $v_t^-=\ell_t(1-\gamma_t^-)$, and $\gamma_t^+$ selects the discharging limit $v_t^+=u_t\gamma_t^+$. The third coefficient places the baseline within this range: $b_t=v_t^-+(v_t^+-v_t^-)\gamma_t^{\mathrm b}$. The reserve capacities follow as $q_t^+=v_t^+-b_t$ and $q_t^-=b_t-v_t^-$. The committed range includes zero, and either power limit can be sustained for one interval in every scenario.

The response head uses a multilayer perceptron (MLP) to produce
a scalar adjustment $\epsilon_{t,s}$ from the shared summary,
observed demand history, battery state, commitments, and interval index. The response decoder adjusts the power needed to maintain the previous grid import,
$\widehat P_{t,s}^{\mathrm{DC}}-g_{t-1,s}$, with $g_{0,s}=g_0$. The learned adjustment allows the response to balance smoothing with available energy and scheduling costs. For $v_t^+>v_t^-$, the reference position within the committed range is $r_{t,s}=(\widehat P_{t,s}^{\mathrm{DC}}-g_{t-1,s}-v_t^-)/(v_t^+-v_t^-)$, clipped strictly between zero and one before applying the logit. The proposed net BESS power is
\begin{equation}\label{eq:response_proposal}
\widetilde p_{t,s}^{\mathrm B}=v_t^-+(v_t^+-v_t^-)
\operatorname{sigmoid}\!\left(\operatorname{logit}(r_{t,s})+\epsilon_{t,s}\right).
\end{equation}
If the two power limits coincide, the output equals that common value. The decoder then clips the proposed power to the part of the committed range that satisfies the power, energy, and non-export limits: $p_{t,s}^{\mathrm B}=\operatorname{clip}_{[L_{t,s},U_{t,s}]}(\widetilde p_{t,s}^{\mathrm B})$. This net power determines the reserve activation $a_{t,s}=p_{t,s}^{\mathrm B}-b_t$, charging power $p_{t,s}^{\mathrm{ch}}=[-p_{t,s}^{\mathrm B}]_+$, and discharging power $p_{t,s}^{\mathrm{dis}}=[p_{t,s}^{\mathrm B}]_+$, where $[p]_+=\max\{p,0\}$. This split prevents simultaneous charging and discharging, with $z_{t,s}=1$ for positive net power and zero otherwise. The remaining variables in $\mathbf x^{\mathrm{rec}}$ follow directly: energy and grid import are computed using~\eqref{eq:soc_dynamics} and~\eqref{eq:grid_power}, and $\delta_{t,s}^{\pm}$ record the ramp changes exceeding the limits in~\eqref{eq:grid_ramp_limits}. The updated energy levels determine the next interval's commitment bounds.

\subsection{Training}
Training first teaches BS-NOL to generate battery schedules, then teaches the predictor to improve the decisions made through this layer. Let $F(\mathbf x;\mathbf H,\boldsymbol{\sigma})$ be the objective in~\eqref{eq:stochastic_model} with the CVaR auxiliary variables minimized out. Here, $\mathbf x=(\mathbf x^{(1)},\mathbf x^{\mathrm{rec}})$ contains the scheduling decisions, $\mathbf H=(H_{t,s})_{t\in\mathcal T,s\in\mathcal S}$ contains the selected historical demand profiles, and $\boldsymbol{\sigma}=(E_0,p_0^{\mathrm B},g_0)$ is the initial state, with $p_0^{\mathrm B}$ the initial net BESS power.

For each training instance $i$, BS-NOL with parameters $\omega$
uses historical demand $\mathbf H_i$ and initial state
$\boldsymbol{\sigma}_i$ to generate a schedule
$\widehat{\mathbf x}_{\omega,i}$ of commitments and battery responses.
We train $\omega$ on $N$ instances using
\begin{equation}\label{eq:nol_training}
\mathcal L_{\mathrm{NOL}}(\omega)=\frac{1}{N}\sum_{i=1}^{N}
F\!\left(\widehat{\mathbf x}_{\omega,i};\mathbf H_i,\boldsymbol{\sigma}_i\right).
\end{equation}
The loss jointly trains the encoder and both output heads by evaluating reserve costs, operating costs, and ramping risk under the input demand profiles, without optimal-solution labels.

Next, we freeze BS-NOL at $\omega^\star$ and use forecast-based
scenarios from~\eqref{eq:scenario_prediction} to generate full-horizon
commitments and the shared summary. Keeping both fixed, we recompute
responses under historical demand $\mathbf H_i$ from the same initial
state using the same decoder. This evaluates the commitments when
demand differs from the forecast.

With these commitments fixed, discharging too much now may leave insufficient energy for a later commitment, while charging too much may leave insufficient space for later charging. To meet later commitments, the decoder keeps the energy remaining after interval $t$ within $\mathcal E_t=[\underline E_t,\overline E_t]$, where $\underline E_t$ is the minimum energy needed for later discharging and $\overline E_t$ leaves enough space for later charging. At the end of the planning horizon ($t=T$), no further commitments remain, so $\mathcal E_T=[E_{\min},E_{\max}]$. For demand above the rated discharge power, we compute these ranges backward from $t=T$ to $t=1$:
\begin{equation}\label{eq:backward_energy}
\mathcal E_{t-1}=\mathcal A_t\cap
\left[\underline E_t+\eta_{\mathrm{ch}}\Delta t\,v_t^-,\ \overline E_t+\frac{\Delta t}{\eta_{\mathrm{dis}}}v_t^+\right].
\end{equation}
Here, $\mathcal A_t$ contains the starting energy levels that
support the current commitment under~\eqref{eq:reserve_energy}.
The bracketed interval contains those from which charging or
discharging within the committed limits can bring the battery
into $\mathcal E_t$, following~\eqref{eq:soc_dynamics}.
Their intersection therefore supports both the current and
remaining commitments.

The responses recomputed within these energy ranges, together with
the unchanged forecast-based commitments, form
$\widehat{\mathbf x}_{\theta,i}$.
We train $\theta$ to minimize its cost and ramping risk:
\begin{equation}\label{eq:dfl_training}
\mathcal L_{\mathrm{DFL}}(\theta)=\frac{1}{N}\sum_{i=1}^{N}
F\!\left(\widehat{\mathbf x}_{\theta,i};\mathbf H_i,\boldsymbol{\sigma}_i\right).
\end{equation}
Gradients pass through the frozen BS-NOL and piecewise differentiable decoder to update only $\theta$. After training, both networks generate decisions through forward passes with fixed parameters and no optimization solver.

\section{Simulation and Discussion}\label{Simulation and Discussions}

\subsection{Experimental Setup}
We use AI facility profiles (10 MW, 2,840 nodes, 80\% target occupancy) from~\cite{vercellino2026dataset}. Chronological splits yield 4,500 training and 128 validation instances. Simulation spans 10,000 one-minute decisions with 60-min horizons. Under a discrete-time assumption, each interval's average demand is known after commitment and before dispatch. The 1-MW/1-MWh BESS has 10--90\% SOC limits, 50\% initial/target SOC, and charging/discharging efficiencies of 0.95/0.90. Illustrative smoothing-focused costs are $\pi_t=80$, $c_t^\pm=20$, $c^{\mathrm{deg}}=2$, and $c^{\mathrm{term}}=100$ in weighted cost units (CU)/MWh, with $R^\uparrow=R^\downarrow=0$, $\lambda^\pm=1$ CU/kW, $\alpha=0.7$, and $\beta=0.5$. All predictors use a $32\!\rightarrow\!32\!\rightarrow\!60$ MLP. MSE+Gurobi and MSE+BS-NOL pair MSE-trained forecasts with Gurobi and frozen BS-NOL, respectively, where the latter isolates decision-focused predictor training. BS-NOL uses 8 scenarios due to limited historical profiles available during training. MSE+Gurobi uses 32, which performed best among 8, 16, and 32 in preliminary tests. The Gurobi solver (version 13.0.3) has a 60-s solver-time budget per decision. Adam settings are batch size 16, early-stopping patience 10, and learning rates $10^{-2}$ for MSE and $10^{-3}$ for BS-NOL/DFL. All experiments run on an Apple M3 Pro CPU.

\subsection{Simulation Results and Discussion}

\begin{figure*}[!h]
\centering
\includegraphics[width=1\linewidth,keepaspectratio]{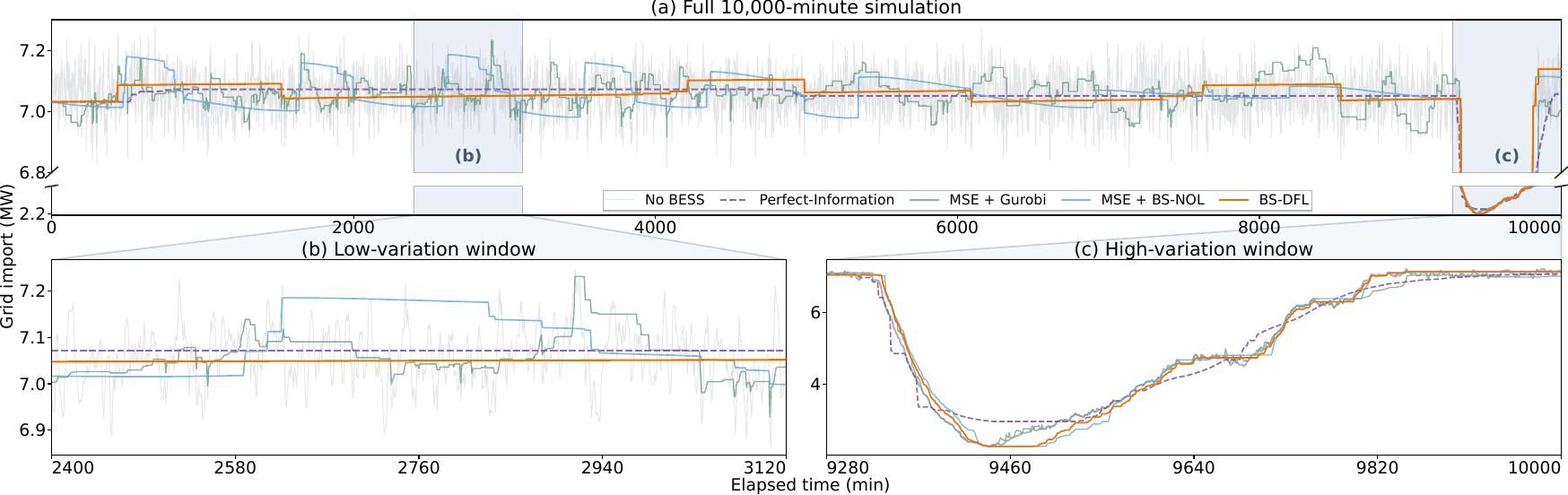}
\caption{Grid-import power over the 10,000-minute evaluation trace. (a) Full trajectory with a broken vertical axis. (b)-(c) 12-hour windows with the lowest and highest mean absolute minute-to-minute demand changes, respectively, shown on different vertical scales.}
\label{fig:simulation}
\end{figure*}

\begin{table}[!h]
\caption{Continuous scheduling performance comparison}
\label{tab:scheduling_results}
\centering
\footnotesize
\setlength{\tabcolsep}{0.35pt}
\renewcommand{\arraystretch}{1.1}
{\fontsize{7.5}{9}\selectfont
\begin{tabular*}{\columnwidth}{@{\extracolsep{\fill}}lrrrrr@{}}
\hline
Metric & No BESS & Perf-Info. & MSE+Gurobi & MSE+BS-NOL & \textbf{BS-DFL} \\
\hline
Mean $|\Delta g|$ & 34.7274 & 0.8237 & 3.5293 & 1.1626 & 0.9988 \\
P95 $|\Delta g|$ & 86.6182 & 0.4066 & 23.4497 & 0.0971 & 0.0104 \\
Maximum $|\Delta g|$ & 199.949 & 986.790 & 413.191 & 179.400 & 203.920 \\
Total cost & 439,557 & 100,832 & 128,382 & 106,500 & 105,257 \\
Reserve cost & 0 & 215 & 717 & 2,485 & 2,898 \\
Forecast MAE & --- & --- & 179.97 & 179.97 & 240.66 \\
\hline
Mean latency & --- & --- & 2,513.41 & 24.76 & 18.77 \\
P95 latency & --- & --- & 5,581.17 & 26.77 & 21.21 \\
Maximum latency & --- & --- & 31,566.21 & 83.51 & 64.07 \\
\hline
\end{tabular*}}
\par\smallskip
\parbox{\columnwidth}{\footnotesize All methods use the same 10,000-minute trace. $\Delta g$ denotes one-minute grid-import change; P95 is the 95th percentile; MAE (mean absolute error) covers 600,000 forecast-truth pairs from rolling 60-step forecasts, including overlapping horizons. Power and MAE: kW; costs: CU; latency: ms. Total cost includes energy, reserve, degradation, ramp, and final SOC penalties, excluding CVaR. Costs are rounded to the nearest CU. Latencies are controller wall times from the respective runs.}
\end{table}

Fig.~\ref{fig:simulation} and Table~\ref{tab:scheduling_results} compare scheduling performance over the 10,000-minute evaluation trace. No BESS denotes uncontrolled demand, while Perfect-Information assumes known future demand within each 60-minute planning horizon. In Fig.~\ref{fig:simulation}(a), BS-DFL exhibits extended periods of nearly constant grid import, resembling the perfect-information reference. The enlarged windows in Fig.~\ref{fig:simulation}(b)-(c) further illustrate its ability to smooth grid import fluctuations under both low- and high-variability conditions. Compared with MSE+Gurobi, BS-DFL reduces mean absolute grid-import changes by 71.70\%. Its P95 change of 0.0104 kW indicates nearly constant grid import during most intervals, although occasional large changes remain, as reflected by the maximum of 203.92 kW. Under the zero-ramp-limit settings, this behavior is consistent with the cumulative absolute change penalty in \eqref{eq:violation_cost}. The largest grid-import changes across the BESS methods occur during the sharp demand decline in Fig.~\ref{fig:simulation}(c). Limited look-ahead may lead to these changes, as the subsequent demand trough and recovery may not yet be visible within the 60-minute prediction horizon.

Despite an additional reserve expenditure of 2,181 CU, BS-DFL reduces total cost by 18.01\% relative to MSE+Gurobi, bringing it within 4.39\% of the perfect-information reference. The resulting reduction in ramp penalties more than offsets the additional reserve expenditure. MSE+BS-NOL improves scheduling performance over MSE+Gurobi through smoother grid import and lower total cost, demonstrating the benefit of the neural optimization layer. BS-DFL further reduces mean absolute changes by 14.09\% and total cost by 1.17\% relative to MSE+BS-NOL, while lowering P95 from 0.0971 to 0.0104 kW. These improvements, despite a higher forecast MAE, demonstrate the value of training predictors directly for downstream scheduling performance. BS-DFL also achieve a mean decision time of 18.77 ms, with a maximum of 64.07 ms, providing computational headroom for more frequent dispatch updates.

\section{Conclusion and Future Work}\label{Conclusion}

This work developed BS-DFL to align demand forecasting with the operating costs and ramping risks of battery scheduling. Compared with an MSE-trained predictor coupled with a MILP solver, BS-DFL reduced mean absolute grid-import changes by 71.70\% and total operating cost by 18.01\%, while achieving an average decision time of 18.77~ms without online optimization. Future work will investigate learned scheduling strategies, electricity-price uncertainty, and robustness across battery configurations and operating conditions.
\bibliographystyle{IEEEtran}
\bibliography{ref}

\end{document}